\documentclass[10pt,conference]{IEEEtran}
\IEEEoverridecommandlockouts
\usepackage{cite}
\usepackage{amsmath,amssymb,amsfonts}
\usepackage{algorithmic}
\usepackage{graphicx}
\usepackage{textcomp}
\usepackage{xcolor}
\usepackage{url,hyperref}
\usepackage{enumitem}

\def\bridges{\textsc{Bridges2023}}
\def\oldbridges{\textsc{Bridges2019}}

\def\BibTeX{{\rm B\kern-.05em{\sc i\kern-.025em b}\kern-.08em
    T\kern-.1667em\lower.7ex\hbox{E}\kern-.125emX}}
\begin{document}

\newcommand{\nb}[2]{
	{
		{\color{red}{
				\fbox{\bfseries\sffamily\scriptsize#1}
				{\sffamily$\triangleright~${\it\sffamily #2}$~\triangleleft$}
	}}}
}
\newcommand\revise[1]{\nb{Revised}{\color{red}#1}}

% SOURCE FOR CODE SNIPPET % https://tex.stackexchange.com/questions/458204/ieeetran-document-class-how-to-align-five-authors-properly/458208#458208
\makeatletter
\newcommand{\linebreakand}{%
  \end{@IEEEauthorhalign}
  \hfill\mbox{}\par
  \mbox{}\hfill\begin{@IEEEauthorhalign}
}
\makeatother

\title{Building \textsc{Bridges} across Papua New Guinea's Digital Divide in Growing the ICT Industry}

\author{\IEEEauthorblockN{Marc Cheong}
\IEEEauthorblockA{\textit{The University of Melbourne} \\
Parkville, Australia \\
marc.cheong@unimelb.edu.au}
\and
\IEEEauthorblockN{Sankwi Abuzo}
\IEEEauthorblockA{\textit{PNG University of Technology} \\
Lae, Papua New Guinea \\
abuzo.sankwi@pnguot.ac.pg}
\and
\IEEEauthorblockN{Hideaki Hata}
\IEEEauthorblockA{\textit{Shinshu University} \\
Nagano, Japan \\
hata@shinshu-u.ac.jp}
\linebreakand

\IEEEauthorblockN{Priscilla Kevin}
\IEEEauthorblockA{\textit{PNG Digital ICT Cluster Inc.} \\
Port Moresby, Papua New Guinea \\
priscilla.kevin@gmail.com}
\and
\IEEEauthorblockN{Winifred Kula}
\IEEEauthorblockA{\textit{eNovaX Solutions} \\
Port Moresby, Papua New Guinea \\
Winifredv.kula@gmail.com}
\and
\IEEEauthorblockN{Benson Mirou}
\IEEEauthorblockA{\textit{PNG University of Technology} \\
Lae, Papua New Guinea \\
benson.mirou@pnguot.ac.pg}
\linebreakand

\IEEEauthorblockN{Christoph Treude}
\IEEEauthorblockA{\textit{Singapore Management University} \\
Singapore, Singapore \\
ctreude@smu.edu.sg}
\and
\IEEEauthorblockN{Dong Wang}
\IEEEauthorblockA{\textit{Tianjin University} \\
Tianjin, China \\
d.wang@ait.kyushu-u.ac.jp}
\and
\IEEEauthorblockN{Raula Gaikovina Kula}
\IEEEauthorblockA{\textit{Osaka University} \\
Osaka, Japan \\
raula-k@ist.osaka-u.ac.jp}
}

\maketitle

\begin{abstract}
Papua New Guinea (PNG) is an emerging tech society with an opportunity to overcome geographic and social boundaries, in order to engage with the global market. However, the current tech landscape, dominated by Big Tech in Silicon Valley and other multinational companies in the Global North, tends to overlook the requirements of emerging economies such as PNG. This is becoming more obvious as issues such as algorithmic bias (in tech product deployments) and the digital divide (as in the case of non-affordable commercial software) are affecting PNG users. The Open Source Software (OSS) movement, based on extant research, is seen as a way to level the playing field in the digitalization and adoption of Information and Communications Technologies (ICTs) in PNG. This perspectives paper documents the outcome of the second \textit{International Workshop on BRIdging the Divides with Globally Engineered Software} (\bridges)~in the hopes of proposing ideas for future research into ICT education, uplifting software engineering (SE) capability, and OSS adoption in promoting a more equitable digital future for PNG. 

\end{abstract}

\begin{IEEEkeywords}
Papua New Guinea, digitalization, digital divide, software development, education, accessibility \end{IEEEkeywords}

\section{Introduction}
As technology benefits from increased market competition, research, and advancements in underlying hardware, end users increasingly benefit from new and improved integrations of such technology in their everyday lives. The best example is the public release of ChatGPT in November 2022, which contributed to the race of consumer-facing applications with Large Language Models (LLMs) -- in the broader class of generative AI systems (GenAI) -- at its core.

Ethicists and scholars on the social implications of technology have come up with the need for oversight and auditing of such products to avoid downstream negative societal impacts.

From the perspectives of technology platforms themselves, using an example from banking, the ``[a]symmetry of value creation occurs when the power, resources, or strategic positioning are unevenly distributed among platform participants'' \cite{Lasak2024-er} which is another form of inequity faced by, say, customers and end-users and their stakeholders.

Specifically in software engineering (SE) and development, there exist ``micro-inequities and barriers'' \cite{Guzman2024-yz} affecting equality and fairness: gender and age inequity \cite{Guzman2024-yz}, geographic and regional bias \cite{Kadji-Ngassam2024-vj}, and potential lack of diversity due to corporate involvement \cite{Newton2024-xu}, are some examples. A {recent} topic of discussion is the deployment of GenAI like the aforementioned ChatGPT: scholars have {similarly} identified concerning biases within such systems, including regional, gender, and racial bias~\cite{cheong2024-dalle,cao2023assessing}. 

However, these discussions are largely found in developed countries, where tech giants have a substantial user base and potential profits in these markets. What about the technology industries, and their user bases, in the Global South? More importantly, how do Software Engineering practitioners, AI researchers, ethicists and social scientists, and policymakers, amongst other stakeholders, begin this conversation for countries where resources are particularly limited? 

Our contribution in this perspectives paper --- stemming from the second \textit{International Workshop on BRIdging the Divides with Globally Engineered Software} (\bridges)\footnote{\url{https://naist-se.github.io/PNG-BRIDGES/}} in Port Moresby, Papua New Guinea --- seeks to discuss several blind spots in efforts to improve software engineering and humanizing its products. 

Specifically, we consider how software engineering and development -- again, in the context of emerging and emergent tech such as GenAI -- often neglect developing nations like Papua New Guinea (PNG), where local voices remain unheard. Notably, this remains an ongoing challenge, as the voices of PNGeans are still not being adequately factored into these discussions.

Furthermore, as Papua New Guinea is the most linguistically-diverse in the world~\cite{The_Economist2017-bc}, and has the most biodiversity of fauna~\cite{UN_Environment_ProgrammeGRID-Geneva_undated-jc}, this potentially is something that the global community can also benefit from. In other words, PNG's engagement in the global Information and Communication Technologies (ICT\footnote{In this paper, we use the terms ICT and IT interchangeably.}) landscape, with emphasis on software and AI, is mutually-beneficial.

This perspectives paper is divided into three parts. Firstly, we provide context on PNG, especially from perspectives of previous studies and workshops on ``challenges for inclusion''~\cite{Kula2022-ey} and inclusivity, specifically in the context of {SE} and digitalization, against the broader landscape of ICT. Secondly, we highlight themes -- especially those raised in \bridges~-- linking to key themes found in extant literature, as well as concerns by PNGean stakeholders. Finally, we propose open research directions in improving software development in PNG, in the broader context of closing the digital divide.

\section{Background}
\label{Background}

\subsection{The Digital Divide}

As reported by Kula and colleagues~\cite{Kula2022-ey} in 2021, based on their findings in \oldbridges, Papua New Guinea (PNG) is characterized as ``...entering the digital age and showing signs of an emerging local software engineering (SE) community''. 

That being said, PNG faces a significant \textit{digital divide}~\cite{sagrista2016}: take the PNGean Internet penetration rate for instance. From $\sim1.3\%$ in 2010~\cite{Kula2022-ey} to $\sim32\%$ as of 2021~\cite{International_Telecommunication_Union_undated-oi}, it is still lower in comparison to its neighbours (say, Fiji, at $\sim84\%$ as of 2018)~\cite{Kula2022-ey}.

The Lowy Institute characterizes the PNGean digital divide succinctly as follows: 

\begin{quote}
    ``Most communities in PNG currently do not have access to digital infrastructure such as broadband [I]nternet. But among those communities that do have access, not everyone has the skills and knowledge to use digital technologies effectively... creat[ing] inequalities and limit[ing] the benefits of digital technology for communities in rural and poorer areas... a digital divide within PNG''~\cite{Mihai_Sora2023-or}. 
\end{quote}

\subsection{Tech Industry: (Lack of) Consideration?}

A major factor which exacerbates the digital divide is that major tech companies like Amazon and Microsoft often overlook the country because their pricing models are out of reach for the average PNGean user. This hinders efforts to leverage ICT for improving living standards and business practices. 

Consider the position of a PNGean entrepreneur, for example. Even basic tasks like invoice printing, flyer creation, and general business administration become expensive due to these prohibitive fees. How can Papua New Guinean entrepreneurs benefit from ever-sophisticated software, when the barriers-to-access are prohibitive, and the software may not be adaptable to the local context?

Also, when it comes to issues such as introducing new AI technologies in the hopes of uplifting PNGeans (ChatGPT, for instance), their voices are still not being heard in the development process of these technologies. It is important to note the rich diversity amongst PNGeans: there are over $\sim840$ language groups alone~\cite{Kula2022-ey, The_Economist2017-bc}. 

Note, however, that Tok Pisin, an official language of PNG, is hardly represented in modern tech products including Google Translate\footnote{Checked on Google Translate as of 17 April 2024 at \url{https://translate.google.com/}.}. Thus, it is crucial to consider questions like: \textit{How many Papua New Guinean elders, scholars, academics, and community leaders are involved in auditing the potential biases within modern AI and software technologies?} As observed by van Reijswoud, many ICT solutions -- largely borrowed from extant Australian counterparts -- ``...fail because they are not geared to the specific needs of Papua New Guinea''~\cite{Van_Reijswoud_Victor2020-rx}.

\subsection{The Role of OSS in PNG}

The use of open-source software (OSS) has, thus, emerged as a potential solution~\cite{Kula2022-ey,Van_Reijswoud_Victor2020-rx}. OSS allows for inclusive access to Information and Communications Technology (ICT) in various sectors like education, business, and government. Proponents of OSS see it as a way forward to bridge the gap between the `haves' and `have-nots' in the digital world.

Examples of popular open-source tools include the Firefox web browser, Linux operating systems, and the LibreOffice suite. These tools are freely accessible under open licenses, allowing users to utilize and adapt them at no cost: removing the barrier of expensive licensing fees compared to closed-source products (such as Microsoft Office, with annual subscription costs per user close to $\sim100$ hours' salary at minimum wage\footnote{An estimate of AUD109.00 for an annual Microsoft 365 license, when compared to an approximate figure of AUD1.00 per hour of minimum wage. See \url{https://www.microsoft.com/en-au/microsoft-365/buy/compare-all-microsoft-365-products} and \url{https://www.lcci.org.pg/wp-content/uploads/2013/03/8-March-2013_2.pdf}.}). 

{Hearteningly, by referring to the statistics on GitHub's Innovation Graph \cite{Github-Inc2024-av}, in Papua New Guinea, there are over 7,547 developers contributing to 6,130 repositories in Q1 2024. Compared to Q1 2020, this is $\sim479\%$ the amount of PNGean developers (1,576 in Q1 2020), and $\sim470\%$ of repositories (1,303 in Q1 2020), in just the span of four years!}

%%%%%
%%%%%

\section{Preliminaries}
\label{KeyThemes}

The theme for \bridges, motivating this perspectives paper, is ``\textit{Pathways for Tech, Big Tech, and Start-up Employment, Competency, and Global Engagement from a Papua New Guinea Perspective}''\footnote{Theme as quoted in the event page: \url{https://naist-se.github.io/PNG-BRIDGES/}}. 

{Spread across multiple days, \bridges~hosted participants from a variety of backgrounds, including industry, academia, not-for-profit (NFP) sector, and government organisations\footnote{\url{https://naist-se.github.io/PNG-BRIDGES/}}. The format is a mix of formal introductions, invited talks in a workshop setting with Q\&A (with 22 participants, discussants, and observers), as well as an invite-only discussion session on selected topics on OSS in SE (Section \ref{DiscussionAndConclusion}).

{As per the Association for the Advancement of Artificial Intelligence (AAAI) reporting guidelines for its workshops,  this perspectives paper on \bridges~sets out to provide ``... historical context, cover highlights, and provide an overview of the topics discussed'' \cite{aaai2023-ic}.  In terms of this paper's treatment of the talks and subsequent discussion, the "methodological frame[s]" employed include "workshops as practice" and "workshops as research methodology" \cite{Orngreen2017-aj}. Ørngreen and Levinsen state that using workshops as a methodology "... aim[s] to produce reliable and valid data about the domain in question regarding forward-oriented processes" \cite{Orngreen2017-aj}, which befits our perspectives paper's outlook.

To promote discussion between academic experts (in SE with a view of OSS, AI, and allied disciplines), governmental, and private sector stakeholders, \bridges~presented four international talks from a diverse array of fields, as a starting point for the consideration of the participants to reflect on the current landscape of technology and OSS, as well as key themes affecting the sector in PNG. 

\begin{description}
    \item[\textbf{Case I. Open Source and Education}] \hfill\\Teaching software development via contributions to real-world OSS projects, showcasing examples, learning outcomes, and soft skills training. This bridges the gap between industry and academic teaching~\cite{salerno2023}.
    \item[\textbf{Case II. Privacy-Preserving Techniques in SE}] \hfill\\This presentation spearheaded discussions on software, privacy, and introducing new techniques to fix extant problems, using the technique of Cross Project Defect Prediction (CPDP) methods for ``a defect prediction model... without sharing any secret code''~\cite{yamomoto2023}. %%
    \item[\textbf{Case III. Diversity and Societal Considerations for AI}] \hfill\\Discussing extant issues in AI-based discrimination, bias, and representational harm. This talk emphasizes GenAI products and concerns about their deployment~\cite{cheong2024-dalle} -- especially if made a component of other software projects -- with a focus on the PNGean context.
    \item[\textbf{Case IV. Managing Community Commons}] \hfill\\Managing the \textit{commons} -- or shared resources -- ranging from natural resources, to software (and OSS) and data, can be tricky. A university experiment on electric vehicle (EV) sharing sheds light on sharing practices, logistics, and patterns~\cite{nakasai2019}.
\end{description}

From the discussion\footnote{{Contributors from \bridges~to the discussion in this paper are credited as the 9 co-authors. However, we also duly acknowledge other members in \bridges~(in the capacity of discussants and organisers) in the Acknowledgements section.}} generated from these four international talks (\textbf{I} --- \textbf{IV}), experts and stakeholders from PNG have debated several themes pertaining to PNGean SE (focusing on OSS), and ICT more broadly.

These themes are set out in the following section.

\section{PNG: Key Themes in SE and ICT}

\subsection{Current State: Software and Higher Education in PNG}
Firstly (viz. Theme \textbf{(A)}), participants agree that software engineering and development training in universities are important, which is a recurring theme in a prior \textsc{Bridges} workshop~\cite{Kula2022-ey}, and was a major theme for discussion in Cases \textbf{I} and \textbf{IV}. However, \textit{contra} the importance, applicability of such techniques and new concepts in industry, participants found that industry is reluctant to share data for academic use.

In PNG, participants note that programming is only taught at the universities and other tertiary institutions. Most students learn programming for the first time when they get into a Computer Science or IT/ICT program (at tertiary level). It takes so much effort for PNG educators to teach them to attain a good proficiency level. Even the general attitude from students is to pass the tests and exams and to continue to meet the requirements to graduate -- without truly appreciating the importance or value of these skills. One way participants have brainstormed is the possibility of linking standards and ethics to industry frameworks such as the Skills Framework for the Information Age (SFIA)\footnote{\url{https://sfia-online.org/en}}.

\bridges~participants agreed that academic courses need to be updated to align with specialisations and capability. In addition, a shared view held by PNG academics is that programming fundamentals should also be included in the ICT curriculum taught in high schools and secondary schools: this might be the case around the world\footnote{See, e.g., for the Australian context: \url{https://www.australiancurriculum.edu.au/f-10-curriculum/general-capabilities/information-and-communication-technology-ict-capability/}.}, but lacking in PNG.

However, with the disruption by emerging technologies, how can we promote this alignment? A key example in recent years is on GenAI (e.g., ChatGPT-based) plagiarism, and its detection, in terms of academic integrity. How can educators rethink assessment in ICT and software development/engineering at tertiary level, in terms of aligning it with academic outcomes and (as before) industry standards, \textit{on top of} the new risk of GenAI affecting academic integrity? 

In light of this discussion (cf. Case \textbf{III}), we take heart in fellow scholars' work in promoting ethical AI education in developing nations, such as the ''integration of ethical AI around research, innovation, and capacity building''~\cite{Nakatumba-Nabende2023} advocated by Nakatumba-Nabende, Suuna, and Bainomugisha, amongst others.

\subsection{Tech Companies, Financial Inclusivity, and the Digital Divide}
The second theme (Theme \textbf{(B)}) mentioned in \bridges~is the yet-to-be-solved digital divide, with regards to the introduction of ever-modern technologies. This has links to Cases \textbf{I}, \textbf{III}, and \textbf{IV} presented in the workshop.

As alluded in Section \ref{Background}, it is hard for users in PNG to acquire off-the-shelf software due to their high cost: PNG can be said to be affected by the digital divide~\cite{Mihai_Sora2023-or,International_Telecommunication_Union_undated-oi,Van_Reijswoud_Victor2020-rx}.

The issue of cost being one of the barriers has been documented in extant work in the prior decades. For instance, James~\cite{JAMES200221} in 2002 posited that any solution to diminish the digital divide needs to:

\begin{quote}
``...incorporate low-cost versions of information technology, rather than products designed for the higher average incomes prevailing in the developed countries.''~\cite{JAMES200221}
\end{quote}

According to participants, Big Tech considers PNG to be outside their target market, resulting in it often being overlooked. Further, consider that only about 13.7\% of the population is considered urban; PNG's population is ``predominantly a rural'' one~\cite{cia2024}.
In \bridges~participants shared various perspectives on how OSS could ameliorate the divide, with salient points covered in Section \ref{DiscussionAndConclusion}.

\subsection{PNG: Local Context, Safety, Standards}
The local context of PNG is often omitted by technology companies when it comes to their promotion, encouragement-of-uptake, and value proposition for their PNG end users and stakeholders (Theme \textbf{(C)}). 

An anecdote shared  by participants in \bridges~is that new tech companies entering the PNG market are not sensitive to local needs, such as using stock imagery of Pacific Island communities \textit{outside} PNG (instead of using PNG communities) when promoting ICTs.

Participants agree that consumers, deployers, and stakeholders need to ask questions such as: \textit{What is the data they collect? Which [local] policies can we apply? How do we benchmark these technologies (in terms of, e.g., safety)?} An example quoted for the latter point is biased facial recognition technology~\cite{Buolamwini2018-pz, Nakatumba-Nabende2023} -- the discourse around this has mostly originated in developed regions, with minimal involvement from local experts.

Another important question raised by participants is  ``\textit{How do we link current PNG standards [to emerging technology]?}'', given that tech is predominantly developed-nation-centric? The case in point involves GenAI systems such as DALLE or Midjourney: participants reckon that PNG locals miss out, as there is a lack of standards/bodies in PNG (\textit{vis-a-vis} the Australian Computer Society's role, in terms of the Australian ICT context), in providing guidance as to ``\textit{What is good?}'' and ``\textit{What `is' AI for good?}'' when it comes to the deployment/use/evaluation of such GenAI in PNG. 

In \bridges, this discussion ties in with the need for the exchange of knowledge --- such as latest techniques in Software Engineering (e.g., Cases \textbf{II}, \textbf{IV}), as well as societal considerations of AI and other ICT products (e.g., Case \textbf{III}) --- in order to empower PNGeans to build on these discussions for the local context.

\subsection{Lack of Software Development in PNG: OSS as an Avenue}
The last theme (Theme \textbf{(D)}) arising from participant discussions is that the PNG ICT and SE sector lacks in-house software development teams. For example, in industry, large companies rely on support for their software needs, but they have no in-house developers.

As mentioned in Section \ref{Background}, and in extant literature~\cite{Van_Reijswoud_Victor2020-rx,Kula2022-ey}, participants agree that OSS could be a valuable tool for PNG (\textit{vis-a-vis} Cases \textbf{I} and \textbf{II}), as it could equip university graduates with relevant skills, connect students with industry opportunities, and provide them with industry-relevant knowledge and policies (e.g.,~\cite{salerno2023,yamomoto2023}). {From the literature, OSS is seen to help participants in developing countries enhance their experiences, motivate them, and help them ``explore promising professional opportunities in terms of employment'' \cite{Kadji-Ngassam2024-vj}.}

Participants are heartened by the fact that companies are already starting to adopt OSS. Some examples include scholarships like Google Summer of Code (GSOC) and Rails Girls' Summer of Code (SOC)\footnote{See \url{https://summerofcode.withgoogle.com/} and \url{https://railsgirlssummerofcode.org/} respectively.}, which promote open-source development. 

Having looked at the key themes from discussion threads by participants in \bridges, against the backdrop of \oldbridges, extant literature on the PNG context (in particular OSS, the digital divide, and ICTs), we conclude the paper with a discussion on open research directions.

%%%%%
%%%%%

\section{Discussion and Conclusion}\label{DiscussionAndConclusion}
In our discussion, we have identified several key themes: starting from \textbf{(A)} the current state of software and its teaching/education in PNG; \textbf{(B)} the current state of Big Tech and its lack of inclusivity; \textbf{(C)} considerations for PNGean local context; and \textbf{(D)} OSS as an avenue to improvement. 

Here, we present several questions for future research, in the spirit of \bridges, based on participant feedback as well of those found in the {literature.} 

Firstly, in the interest of themes \textbf{(A)} and \textbf{(D)}, how can we align the interests of stakeholders, in making the teaching of software engineering (and ICT more broadly) accessible to various PNGeans? Specific {questions for future research can include:}

\begin{enumerate}[label={Q\arabic*},series=rqs]
\item {What is the current state of accessibility of SE and ICT education resources amongst PNGeans, and what skillsets/areas are needed the most?}
\item {How can academia and the private sector help promote ICT literacy and skill-building (through, e.g., short courses and science outreach programs)?}
\item {What strategies and policies can be put in place to upskill/reskill workers for ICT-focused jobs?}
\item {What are pathways to promote training in coding and software development for kids and adults?}
\item {What roles can OSS stakeholders (e.g., sponsors of OSS projects) take -- from scholarships to advocacy -- in helping uplift the Global South's ICT capabilities?}
\end{enumerate}

Secondly, using the affordances provided by OSS, how can we promote software industry skills such as code review, version control, and pull requests (e.g.,~\cite{salerno2023}) in the teaching of software development in PNG: covering themes \textbf{(A)},  \textbf{(C)}, and \textbf{(D)}. Suggestions made in this space include:

\begin{enumerate}[label={Q\arabic*},resume*=rqs]
\item {What projects can be used as a `springboard' into using OSS as an educational project?\\
To this point, an invitation-only discussion session at \bridges identified an initial starting point: projects such as community translations of extant OSS packages (Firefox or VLC) into Tok Pisin, taught as part of a software engineering curriculum.}
\item {What long-term academic programs can be formulated for PNG that incorporate OSS in their syllabus?}
\item {How can the tailoring of existing OSS to the local PNGean context benefit more PNGeans -- especially in ameliorating the digital divide?}
\end{enumerate}

Lastly, how could the global community help promote diversity and inclusion in developing ICT regions, such as PNG [themes \textbf{(B)}, \textbf{(C)}, and \textbf{(D)}]? Open research areas include explorations on {the following:}

\begin{enumerate}[label={Q\arabic*},resume*=rqs]
\item {How can OSS projects, at large, counter the effects of non-inclusivity of commercial software?}
\item {How can OSS projects reverse the underrepresentation of the PNG economy in terms of Big Tech and the global tech landscape?}
\item {What areas in existing `off-the-shelf' software/services/ICTs can lead to non-inclusivity for developing ICT regions?}
\item {How are AI-powered ICTs, including future LLMs, embedding biases that disadvantage areas of the Global South?}
\end{enumerate}

As technology has evolved in the past few decades, it is imperative that PNG is included in the global conversation, to promote a more equitable playing field for all, and to empower PNG to be the next ``Silicon Valley of the Pacific''. {The proposed future research questions above are merely a starting point; but each small step contributes in a meaningful way to level the playing field for the Global South's developing ICT regions.}

{For future iterations of \bridges, we envisage further discussion on practicalities, stakeholder experiences, implementation methodologies, and more case studies: with an emphasis on academic and industry collaborations.}

\section*{Acknowledgment}
Our acknowledgement goes to all participants and organizers of \bridges, in particular {Emily Lea Kula, Kingsley Lea, Takashi Kobayashi, Kenichi Matsumoto, and Lena Korugl}, for their gracious hospitality and collegiality. \textit{Tenk yu tru olgeta}!
We thank the support of PNG Department of ICT, NASFUND, NAIST, and the Papua New Guinea University of Technology to make BRIDGES2023 a reality.

We acknowledge the use of Generative AI -- Google Gemini (Bard) -- in the copyediting process. This is specifically limited to correcting dictation transcripts and free-form meeting notes, using the prompt `\textit{The following is a poorly transcribed set of dictated notes. Correct them for readability.}'.\\

\bibliographystyle{IEEEtran}
\bibliography{bib}

\begin{thebibliography}{10}
\providecommand{\url}[1]{#1}
\csname url@samestyle\endcsname
\providecommand{\newblock}{\relax}
\providecommand{\bibinfo}[2]{#2}
\providecommand{\BIBentrySTDinterwordspacing}{\spaceskip=0pt\relax}
\providecommand{\BIBentryALTinterwordstretchfactor}{4}
\providecommand{\BIBentryALTinterwordspacing}{\spaceskip=\fontdimen2\font plus
\BIBentryALTinterwordstretchfactor\fontdimen3\font minus \fontdimen4\font\relax}
\providecommand{\BIBforeignlanguage}[2]{{%
\expandafter\ifx\csname l@#1\endcsname\relax
\typeout{** WARNING: IEEEtran.bst: No hyphenation pattern has been}%
\typeout{** loaded for the language `#1'. Using the pattern for}%
\typeout{** the default language instead.}%
\else
\language=\csname l@#1\endcsname
\fi
#2}}
\providecommand{\BIBdecl}{\relax}
\BIBdecl

\bibitem{Lasak2024-er}
\BIBentryALTinterwordspacing
P.~Łasak and S.~Wyciślak, ``\BIBforeignlanguage{en}{The dichotomy of inclusiveness and vulnerability as a consequence of banking platform development},'' \emph{\BIBforeignlanguage{en}{Research in international business and finance}}, vol.~72, no. 102536, p. 102536, Oct. 2024. [Online]. Available: \url{http://dx.doi.org/10.1016/j.ribaf.2024.102536}
\BIBentrySTDinterwordspacing

\bibitem{Guzman2024-yz}
\BIBentryALTinterwordspacing
E.~Guzmán, R.~A.-L. Fischer, and J.~Kok, ``\BIBforeignlanguage{en}{Mind the gap: gender, micro-inequities and barriers in software development},'' \emph{\BIBforeignlanguage{en}{Empirical Software Engineer}}, vol.~29, no.~1, Jan. 2024. [Online]. Available: \url{http://dx.doi.org/10.1007/s10664-023-10379-8}
\BIBentrySTDinterwordspacing

\bibitem{Kadji-Ngassam2024-vj}
\BIBentryALTinterwordspacing
M.~Kadji~Ngassam, ``\BIBforeignlanguage{en}{Trust and involvement of cameroonian software developers in open-source projects},'' \emph{\BIBforeignlanguage{en}{Data Science and Management}}, vol.~7, no.~4, pp. 332--339, Dec. 2024. [Online]. Available: \url{http://dx.doi.org/10.1016/j.dsm.2024.04.005}
\BIBentrySTDinterwordspacing

\bibitem{Newton2024-xu}
\BIBentryALTinterwordspacing
O.~B. Newton and S.~M. Fiore, ``\BIBforeignlanguage{en}{Understanding participation and corporatization in service of diversity in free/libre and open source software development projects},'' \emph{\BIBforeignlanguage{en}{The Journal of systems and software}}, vol. 217, no. 112163, p. 112163, Nov. 2024. [Online]. Available: \url{http://dx.doi.org/10.1016/j.jss.2024.112163}
\BIBentrySTDinterwordspacing

\bibitem{cheong2024-dalle}
\BIBentryALTinterwordspacing
M.~Cheong, E.~Abedin, M.~Ferreira, R.~Reimann, S.~Chalson, P.~Robinson, J.~Byrne, L.~Ruppanner, M.~Alfano, and C.~Klein, ``Investigating gender and racial biases in dall-e mini images,'' \emph{ACM J. Responsib. Comput.}, mar 2024, just Accepted. [Online]. Available: \url{https://doi.org/10.1145/3649883}
\BIBentrySTDinterwordspacing

\bibitem{cao2023assessing}
Y.~Cao, L.~Zhou, S.~Lee, L.~Cabello, M.~Chen, and D.~Hershcovich, ``Assessing cross-cultural alignment between chatgpt and human societies: An empirical study,'' \emph{arXiv preprint arXiv:2303.17466}, 2023.

\bibitem{The_Economist2017-bc}
\BIBentryALTinterwordspacing
{The Economist}, ``\BIBforeignlanguage{en}{Papua new guinea's incredible linguistic diversity},'' \emph{\BIBforeignlanguage{en}{The Economist}}, Jul. 2017. [Online]. Available: \url{https://www.economist.com/the-economist-explains/2017/07/20/papua-new-guineas-incredible-linguistic-diversity}
\BIBentrySTDinterwordspacing

\bibitem{UN_Environment_ProgrammeGRID-Geneva_undated-jc}
\BIBentryALTinterwordspacing
{UN Environment Programme/GRID-Geneva}, P.~Peduzzi, H.~Dao, A.~De~Bono, P.~Lacroix, T.~Piller, A.~Benvenuti, C.~Gampert, and F.~Mose, ``\BIBforeignlanguage{en}{Biodiversity / papua new guinea},'' \url{https://dicf.unepgrid.ch/papua-new-guinea/biodiversity}, accessed: 2024-5-24. [Online]. Available: \url{https://dicf.unepgrid.ch/papua-new-guinea/biodiversity}
\BIBentrySTDinterwordspacing

\bibitem{Kula2022-ey}
\BIBentryALTinterwordspacing
R.~G. Kula, C.~Treude, H.~Hata, S.~Baltes, I.~Steinmacher, M.~A. Gerosa, and W.~K. Amini, ``Challenges for inclusion in software engineering: The case of the emerging papua new guinean society,'' \emph{IEEE Softw.}, vol.~39, no.~3, pp. 67--76, May 2022. [Online]. Available: \url{http://dx.doi.org/10.1109/MS.2021.3098116}
\BIBentrySTDinterwordspacing

\bibitem{sagrista2016}
\BIBentryALTinterwordspacing
M.~Sagrista and P.~Matbob, \emph{Pacific Journalism Review}, vol.~22, no.~2, p. 20–34, 2016. [Online]. Available: \url{https://search.informit.org/doi/10.3316/informit.598619911837788}
\BIBentrySTDinterwordspacing

\bibitem{International_Telecommunication_Union_undated-oi}
\BIBentryALTinterwordspacing
{International Telecommunication Union}, ``Individuals using the internet (\% of population) - papua new guinea,'' \url{https://data.worldbank.org/indicator/IT.NET.USER.ZS?locations=PG}, accessed: 2024-4-16. [Online]. Available: \url{https://data.worldbank.org/indicator/IT.NET.USER.ZS?locations=PG}
\BIBentrySTDinterwordspacing

\bibitem{Mihai_Sora2023-or}
\BIBentryALTinterwordspacing
{Mihai Sora}, ``\BIBforeignlanguage{en}{Building the {Australia-PNG} digital ecosystem},'' Lowy Institute, Tech. Rep., Jun. 2023. [Online]. Available: \url{https://www.lowyinstitute.org/publications/building-australia-png-digital-ecosystem}
\BIBentrySTDinterwordspacing

\bibitem{Van_Reijswoud_Victor2020-rx}
\BIBentryALTinterwordspacing
{van Reijswoud Victor}, ``The power to change: Adopting free and open source software in papua new guinea,'' \emph{Contemporary PNG Studies}, vol.~10, pp. 40--62, Aug. 2020. [Online]. Available: \url{https://doi.org/10.3316/informit.962921356019753}
\BIBentrySTDinterwordspacing

\bibitem{Github-Inc2024-av}
\BIBentryALTinterwordspacing
{Github, Inc}, ``\BIBforeignlanguage{en}{{PG} | {GitHub} innovation graph},'' 2024. [Online]. Available: \url{https://innovationgraph.github.com/economies/pg}
\BIBentrySTDinterwordspacing

\bibitem{aaai2023-ic}
{Association for the Advancement of Artificial Intelligence}, ``\BIBforeignlanguage{en}{Symposia \& workshop report guidelines},'' \url{https://aaai.org/aaai-publications/contribute/symposia-workshop-report-guidelines/}, Oct. 2023, accessed: 2024-12-23.

\bibitem{Orngreen2017-aj}
\BIBentryALTinterwordspacing
R.~Ørngreen and K.~Levinsen, ``Workshops as a research methodology,'' \emph{The Electronic Journal of eLearning}, vol.~15, no.~1, pp. 70--81, 2017. [Online]. Available: \url{https://files.eric.ed.gov/fulltext/EJ1140102.pdf}
\BIBentrySTDinterwordspacing

\bibitem{salerno2023}
\BIBentryALTinterwordspacing
L.~Salerno, S.~de~Fran\c{c}a Tonh\~{a}o, I.~Steinmacher, and C.~Treude, ``Barriers and self-efficacy: A large-scale study on the impact of oss courses on student perceptions,'' in \emph{Proceedings of the 2023 Conference on Innovation and Technology in Computer Science Education V. 1}.\hskip 1em plus 0.5em minus 0.4em\relax Association for Computing Machinery, 2023, p. 320–326. [Online]. Available: \url{https://doi.org/10.1145/3587102.3588789}
\BIBentrySTDinterwordspacing

\bibitem{yamomoto2023}
H.~Yamamoto, D.~Wang, G.~Rajbahadur, M.~Kondo, Y.~Kamei, and N.~Ubayashi, ``Towards privacy preserving cross project defect prediction with federated learning,'' in \emph{Proceedings - 2023 IEEE International Conference on Software Analysis, Evolution and Reengineering, SANER 2023}, T.~Zhang, X.~Xia, and N.~Novielli, Eds.\hskip 1em plus 0.5em minus 0.4em\relax IEEE, 2023, pp. 485--496.

\bibitem{nakasai2019}
K.~Nakasai, Y.~Ikutani, D.~Takata, H.~Hata, and K.~Matsumoto, ``Toward sustainable communities with a community currency – a study in car sharing,'' in \emph{2019 20th IEEE/ACIS International Conference on Software Engineering, Artificial Intelligence, Networking and Parallel/Distributed Computing (SNPD)}, 2019, pp. 478--483.

\bibitem{Nakatumba-Nabende2023}
\BIBentryALTinterwordspacing
J.~Nakatumba-Nabende, C.~Suuna, and E.~Bainomugisha, \emph{AI Ethics in Higher Education: Research Experiences from Practical Development and Deployment of AI Systems}.\hskip 1em plus 0.5em minus 0.4em\relax Cham: Springer International Publishing, 2023, pp. 39--55. [Online]. Available: \url{https://doi.org/10.1007/978-3-031-23035-6_4}
\BIBentrySTDinterwordspacing

\bibitem{JAMES200221}
\BIBentryALTinterwordspacing
J.~James, ``Low-cost information technology in developing countries: current opportunities and emerging possibilities,'' \emph{Habitat International}, vol.~26, no.~1, pp. 21--31, 2002. [Online]. Available: \url{https://www.sciencedirect.com/science/article/pii/S0197397501000303}
\BIBentrySTDinterwordspacing

\bibitem{cia2024}
\BIBentryALTinterwordspacing
{Central Intelligence Agency}, ``Papua new guinea,'' \url{https://www.cia.gov/the-world-factbook/countries/papua-new-guinea/}, 2024, accessed: 2024-5-14. [Online]. Available: \url{https://www.cia.gov/the-world-factbook/countries/papua-new-guinea/}
\BIBentrySTDinterwordspacing

\bibitem{Buolamwini2018-pz}
\BIBentryALTinterwordspacing
J.~Buolamwini and T.~Gebru, ``Gender shades: Intersectional accuracy disparities in commercial gender classification,'' in \emph{Proceedings of the 1st Conference on Fairness, Accountability and Transparency}, ser. Proceedings of Machine Learning Research, S.~A. Friedler and C.~Wilson, Eds., vol.~81.\hskip 1em plus 0.5em minus 0.4em\relax PMLR, 2018, pp. 77--91. [Online]. Available: \url{https://proceedings.mlr.press/v81/buolamwini18a.html}
\BIBentrySTDinterwordspacing

\end{thebibliography}

\end{document}